\documentclass[11pt]{article}
\usepackage{xspace}
\usepackage{graphicx}
\usepackage{amsmath}
\usepackage{amssymb}
\usepackage{color}

\usepackage{enumitem}
\usepackage{url}
\usepackage{wrapfig}

\definecolor{Red}{rgb}{1,0,0}
\definecolor{Green}{rgb}{0,1,0}
\definecolor{Blue}{rgb}{0,0,1}
\definecolor{Black}{rgb}{0,0,0}

\def\beq{\begin{equation}}
\def\eeq#1{\label{#1}\end{equation}}
\def\eeqn{\end{equation}}

\def\beqa{\begin{eqnarray}}
\def\eeqa#1{\label{#1}\end{eqnarray}}
\def\eeqan{\end{eqnarray}}

\let\bar=\overbar

\def\Dslash{\not{\hbox{\kern-4pt $D$}}}
\def\dslash{\not{\hbox{\kern-2pt $\del$}}}

\def\msb{{\bar{\ssstyle M \kern -1pt S}}}

\def\Title#1{\begin{center} {\Large {\bf #1} } \end{center}}

\begin{document}

\Title{Baryogenesis \\ \medspace {\small A small review of the big picture}}

\bigskip\bigskip


\begin{raggedright}  

Csaba Bal\'azs\index{Balazs, C.}, {\it ARC Centre of Excellence for Particle Physics at the Tera-scale \\ ~~~ School of Physics, Monash University, Melbourne, Victoria 3800, Australia}\\

\begin{center}\emph{I schematically, and very lightly, review some ideas that fuel model building in the field of baryogenesis.  Due to limitations of space, and my expertise, this review is incomplete and biased toward particle physics, especially supersymmetry.}\end{center}
\bigskip
\end{raggedright}

{\small
\begin{flushleft}
\emph{To appear in the proceedings of the Interplay between Particle and Astroparticle Physics workshop, 18 -- 22 August, 2014, held at Queen Mary University of London, UK.}
\end{flushleft}
}

\section{Introduction}

From the point of view of contemporary physics, the Universe is a strange place because it contains a considerable amount of matter.  We are so used to the existence of matter around, or for that matter inside, us that we take it for granted.  We are, however, hard pressed to explain based on known fundamental principles why the Universe contains mostly matter but hardly any antimatter.  
Our cardinal principles of the corpuscular world are encapsulated in the Standard Model (SM) of elementary particles.  This model contains twelve types of matter particles: six quarks and six leptons.  These matter particles are differentiated by two quantum numbers: quarks carry a baryon number and leptons a unit of lepton number.  They all have antimatter partners.  The mass, and all other quantum numbers, of the anti-particles are the same as their partners', with the exception of electric charge, which is opposite.  

Baryogenesis models attempt to understand the mechanism through which the cosmic matter-antimatter asymmetry arises in the framework of elementary particle physics.  They offer mechanisms for creating matter-antimatter asymmetry from an initially symmetric Universe \cite{Rubakov:1996vz}-\nocite{Yajnik:2006kn, Buchmuller:2007fd, Dolgov:2009py, Mazumdar:2011zd, Morrissey:2012db, Boucenna:2013wba, Agashe:2014kda}\cite{Racker:2014yfa}.\footnote{In a similar vein leptogenesis models transfer an asymmetry created in the leptonic sector to todays baryons \cite{Canetti:2012zc}.}
In a remarkable 1967 paper Sakharov established that baryogenesis requires three necessary ingredients \cite{Sakharov:1967dj}: 
\begin{itemize}[noitemsep,topsep=0pt]
 \item baryon number ($B$) violation,
 \item violation of particle-antiparticle ($C$) symmetry, and the combined $C$ and left-right or parity ($CP$) symmetry, and
 \item departure from thermal equilibrium.
\end{itemize}
These conditions are required components of all baryogenesis models.


\section{Prerequisites}

Baryogenesis is a complicated subject and it is impossible to do justice to it on a few pages, let alone in a conference talk.  One reason for its complexity is that baryogenesis is the subject of intensive research and there are almost as many open questions as answers in the field.  Another source of difficulties is that studying baryogenesis requires an intimate knowledge of various sub-fields far beyond elementary particle theory.  
Quantum field theory is an obvious prerequisite, since baryogenesis models are trying to explain the observed cosmic matter-antimatter asymmetry based on fundamental microscopic physics.  Ordinary `zero temperature', or temperature independent, quantum field theory is inadequate to study baryogenesis since the process might be intimately related to a thermal phase transition.  The latter is typically quantified by thermal quantities, such as an order parameter and temperature dependent potential.  So various baryogenesis scenarios require elements of the finite temperature field theory framework.  

Even within the boundaries of zero temperature quantum field theory, a wide range of concepts enter into baryogenesis models.  Classical solutions of quantum field theory, such as instantons and sphalerons, are used to catalyze baryon asymmetry.  The violation of $B$, $C$, $CP$ discrete symmetries is at the heart of the matter.  Anomalies, quantum violations of classical symmetries, are essential for the breaking of $B$.   Fundamental continuous symmetries such as Lorentz and gauge invariance, are mandatory to respect.  Other symmetries, such as scale invariance, may be related to baryogenesis.  The knowledge of the SM of elementary particles is the starting point of investigation.  Since known mechanisms are not able to generate enough matter-antimatter asymmetry, extensions of the SM are required.  Supersymmetric models have been shown to cope with the problem, as well as grand unified theories, or models with right handed neutrinos, or other exotic particles such as Q-balls.  Various models of generating neutrino mass, especially the ones that utilize right handed neutrinos and the see-saw mechanism, play an important role in leptogenesis.    

Beyond quantum field theory, understanding baryogenesis requires various concepts from thermodynamics: the theory of phase transitions and diffusion, thermodynamic fluctuations, kinetic theory, and the list goes on.  Since baryogenesis took place in the early Universe, cosmological issues such as inflation, preheating, or reheating frequently enter into its description.  Astrophysical and particle astrophysical observation, in turn, provides constraints on baryogenesis and the related cosmological questions.  Colliders can also give important bounds on low scale baryogenesis models, thus understanding of some collider phenomenology comes in handy.  Low energy experiments, such as measurements of electric dipole moments and $CP$-violating rare decays also provide important constraints.

\section{The observed baryon asymmetry}

\begin{figure}[!ht]
	\begin{center}
		\includegraphics[width=0.53\columnwidth]{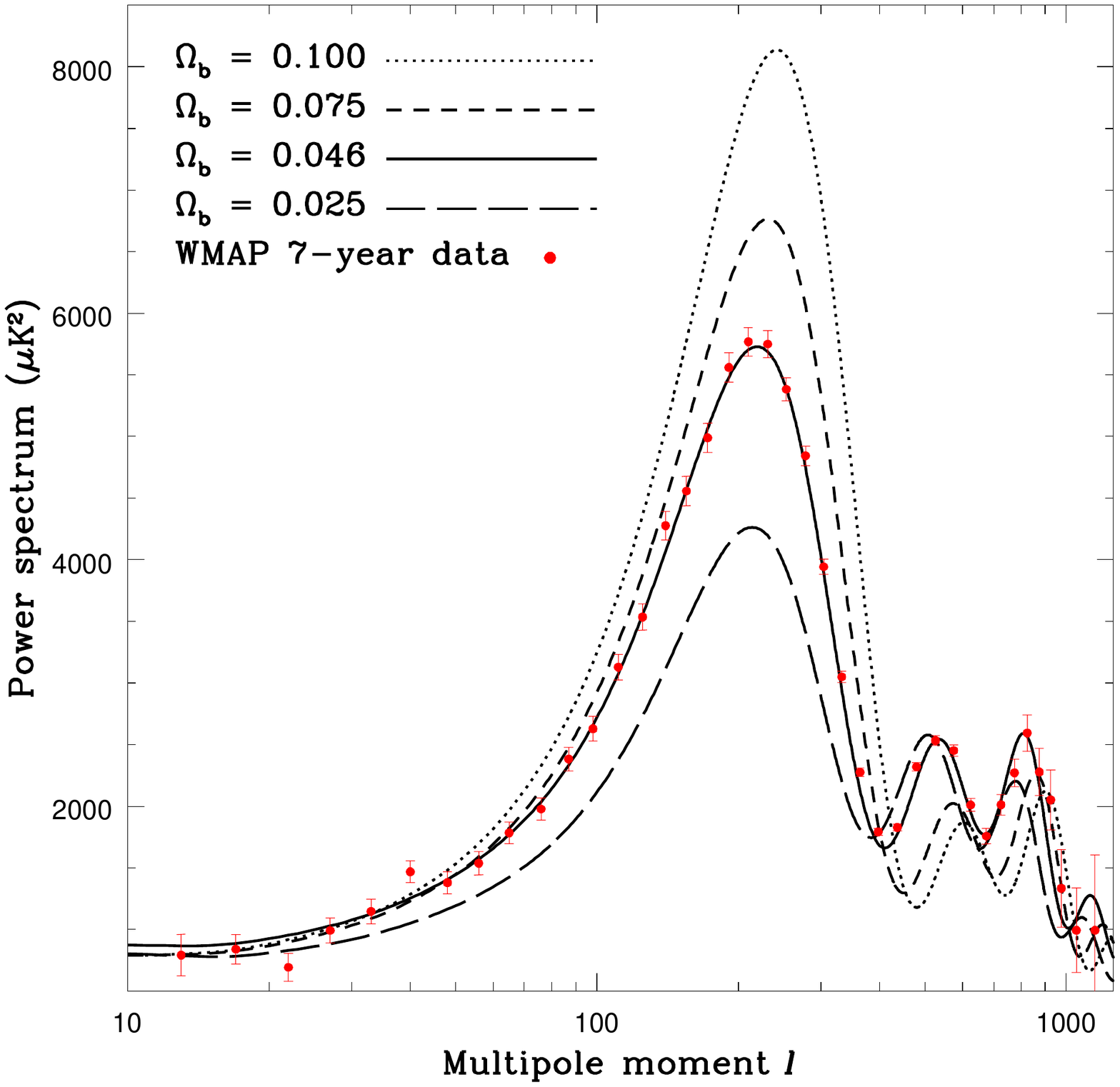} ~~~~~
		\includegraphics[width=0.41\columnwidth]{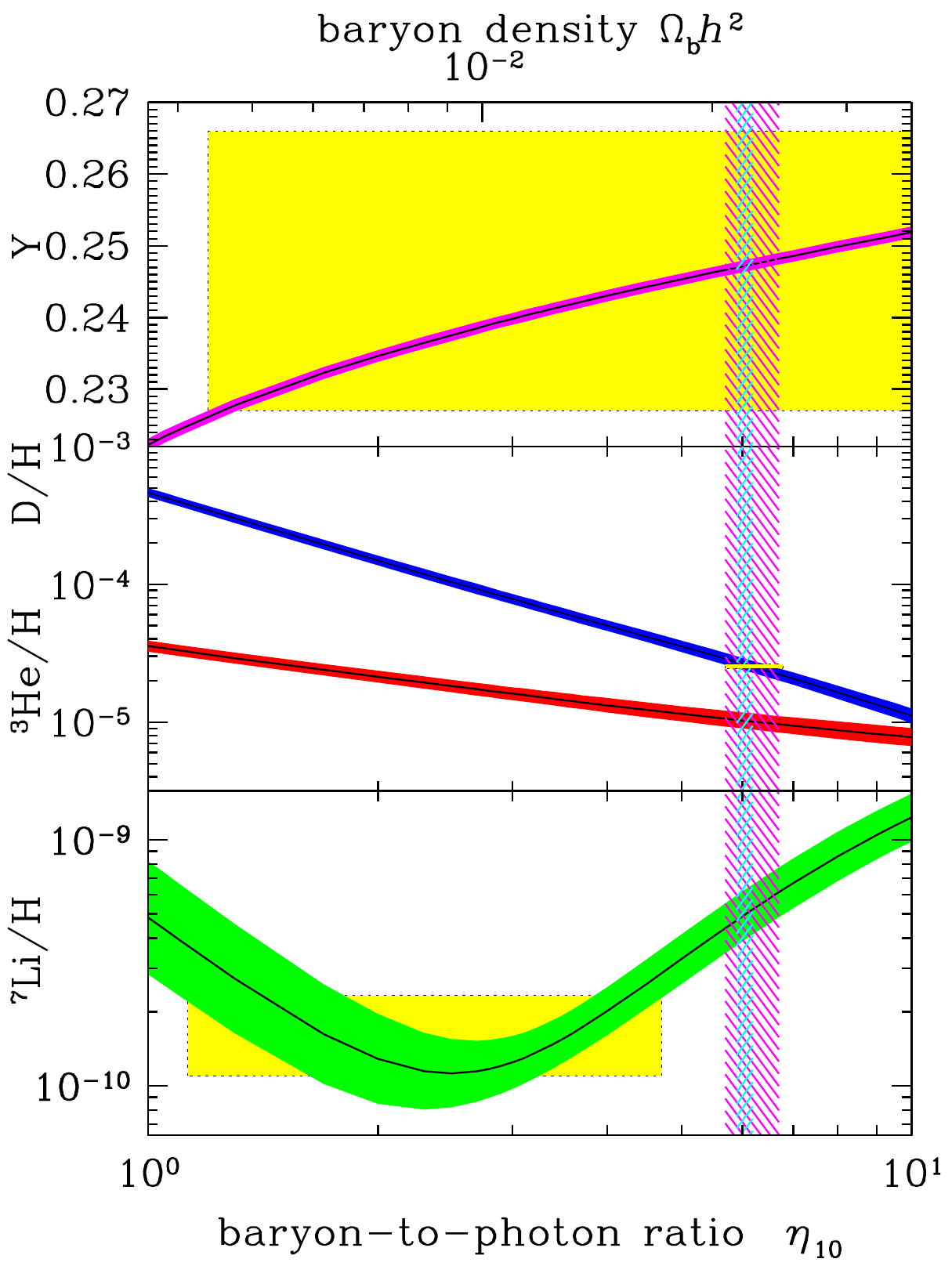}
		\caption{Sensitivity of the first acoustic peak in the Cosmic Microwave Background to the average baryon density of the Universe (left frame).  Concordance between the cosmic abundance of the lightest elements and the baryon density as required by Big Bang Nucleosynthesis (right frame). (Taken from \cite{Garrett:2010hd} and \cite{Agashe:2014kda}).}
		\label{fig:CMBBBN}
	\end{center}
\end{figure}

Precise temperature measurements of the Cosmic Microwave Background (CMB) constrain the total baryon content of the Universe.  As illustrated by the left frame of Figure \ref{fig:CMBBBN}, the first acoustic peak of the CMB is especially sensitive to the amount of baryons in the Universe about 400 thousand years after the Big Bang.  The latest CMB measurements by the Planck satellite constrain the average energy density of t baryons in he Universe to \cite{Ade:2013zuv}
\begin{equation}
\Omega_b = 0.0490 \pm 0.0007 .
\end{equation}
Since, in the units of the critical density, the total energy density is determined to be $\Omega_{total} = 1.00 \pm 0.03$, the above $\Omega_b$ value implies that the Universe, in average, contains about 5 percent baryons \cite{Ade:2013zuv}.  This result is consistent with the amount of baryons required by the observed abundances of the lightest elements and the predictions of Big Bang Nucleosynthesis \cite{Agashe:2014kda} 
\begin{equation}
0.046 \leq \Omega_b \leq 0.056 ~~~ (95\%~{\rm CL}),
\label{eq:Omega_b_BBN}
\end{equation}
as shown in the right frame of Figure \ref{fig:CMBBBN}.

\begin{figure}[!ht]
	\begin{center}
		\includegraphics[width=0.49\columnwidth]{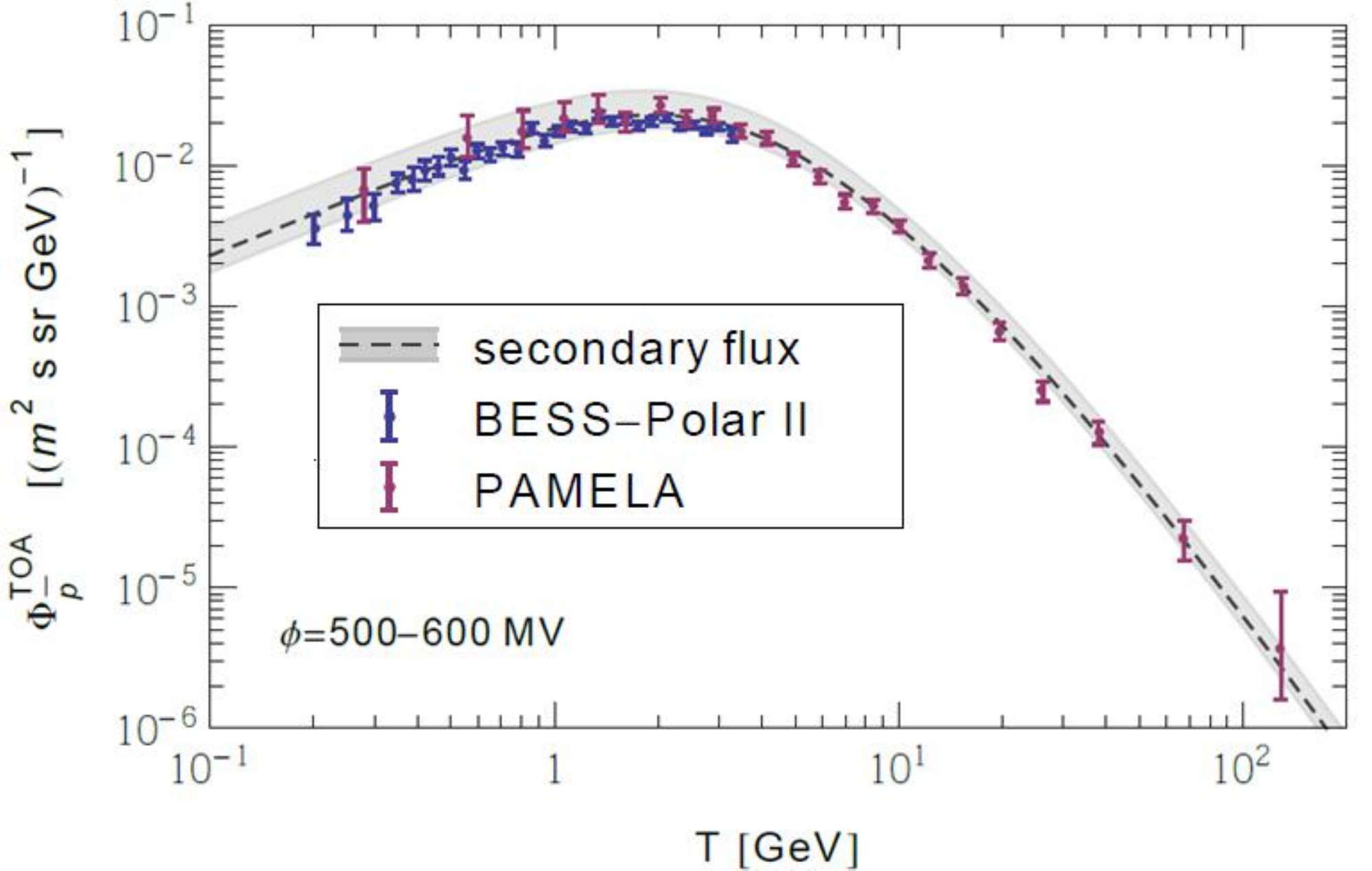}
		\includegraphics[width=0.46\columnwidth]{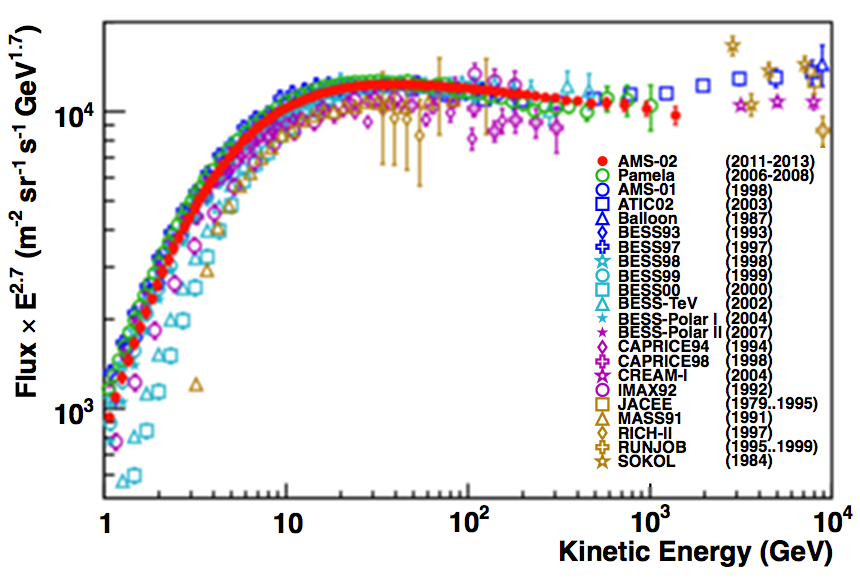}
		\caption{Antiproton cosmic ray flux as measured by AMS-02 and other experimental collaborations.  The observed antiproton flux is consistent with secondary production that is due to collisions of baryons, leptons or photons in the interstellar medium (taken from \cite{Kappl:2014hha, Consolandi:2014uia}).}
		\label{fig:AMS}
	\end{center}
\end{figure}

These numbers are striking because at the same time traces of antibaryons are only found in the cosmic ray data.  As shown in Figure \ref{fig:AMS}, the measurements of the AMS-02 and other  collaborations find the amount of antibaryons consistent with secondary production that is due to collisions of baryons, leptons or photons in the interstellar medium \cite{Kappl:2014hha}.  
The amount of asymmetry measured between baryons and antibaryons is customarily expressed in therms of the total entropy density, or the average photon (number) density $n_\gamma$ as
\begin{equation}
\eta = \frac{n_b - n_{\bar b}}{n_\gamma} .
\label{eq:eta}
\end{equation}
Since the baryonic energy and number densities are related, Eqs.(\ref{eq:Omega_b_BBN}-\ref{eq:eta}) imply that the baryon-to-photon ratio is constrained by Big Bang Nucleosynthesis
\begin{equation}
5.7 \times 10^{−10} \leq \eta \leq 6.7 \times 10^{−10} .
\end{equation}

The puzzle is the following.  It is most likely that shortly after the Big Bang the Universe underwent a radiation dominated period, when the baryon density must have been much lower than presently.  If inflation took place before this period it certainly diluted any preexisting baryon asymmetry.  It is also known that in observed particle collisions baryons and antibaryons are produced in pairs.  So why do we detect much more baryons than antibaryons today?

\section{Baryogenesis in the Standard Model}

Remarkably, the Standard Model of particles offers an explanation for the existence of a baryon asymmetry starting from a baryon-antibaryon symmetric early Universe.  This is because Sakharov's conditions are satisfied in the SM.

\begin{wrapfigure}{r}{0.5\textwidth}
	\begin{center}
		\includegraphics[width=0.35\columnwidth]{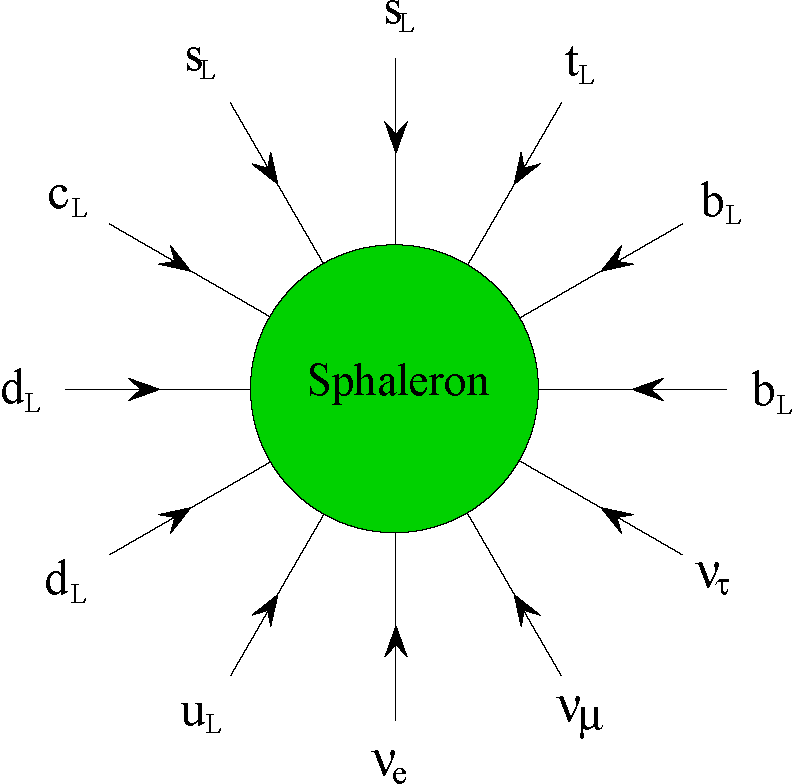}
		\caption{Graphical representation of the Stadard Model electroweak process, referred to as a sphaleron.  The simplest type of sphaleron, pictured, changes baryon number by 3 units (taken from \cite{Buchmuller:2012tv}).}
		\label{fig:sphaleron}
	\end{center}
\end{wrapfigure}

Baryon number conservation is violated in the SM.  More precisely, while $B-L$ is conserved, $B+L$ is anomalous at transitions between inequivalent $SU(2)_L$ vacua.  The SM Lagrangian is invariant under $SU(2)_L$ gauge transformations.  These transformations, however, change the $B+L$ number of the vacuum state of the quantized theory by \cite{'tHooft:1976up, 'tHooft:1976fv}
\begin{equation}
\Delta (B+L) = 2 N_f N_{CS} .
\end{equation}
Here $N_f$ is the number of matter families (3 in the SM), and $N_{CS}$ is the Chern-Simons number which measures the amount the phases of quantum fields have changed under the gauge transformation (in units of $2 \pi$).  Since $B-L$ is conserved, the change in $B$ due to gauge transformations accounts for 
\begin{equation}
\Delta B = \frac{1}{2} \Delta (B+L) = N_f N_{CS} .
\end{equation}
This equation implies that in the SM processes with $\Delta (B+L) = 3, 6, 9 ...$ are possible.  A Feynman diagram-like representation of such a non-perturbative process, simply referred to as a sphaleron, is shown in Figure \ref{fig:sphaleron}.  It has been shown that during the electroweak phase transition these sphalerons can induce a baryon asymmetry \cite{Kuzmin:1985mm}.  

The SM also violates charge conjugation and $C$ combined with parity, $CP$.  It is a chiral theory, which means that left and right handed fermions, defined by the projections $\psi_{L,R} = ((1 \mp \gamma_5)/2) \psi$, behave differently in it.  Only the left handed components of the fermions engage in electroweak interactions.  This feature violates parity maximally.  Since, without quark mixing, $CP$ is conserved in the SM, the above also implies violation of $C$.  A small amount of explicit $CP$ violation is also introduced in the SM via the Cabibbo-Kobayashi-Maskawa (CKM) matrix that mixes down type quarks.

Departure from thermal equilibrium is a notion which goes beyond the temperature independent quantum field theory framework.  The standard cosmology model, $\Lambda$CDM, supplies just this condition because the expansion of space creates out-of-equilibrium conditions.  While this effect is minute, a much more substantial contribution comes from the thermal phase transition associated with the electroweak symmetry breaking.  During a first order electroweak phase transition bubbles of the broken vacuum form in an unbroken phase.  The expansion of these bubble walls is shown to lead to considerable departure from thermal equilibrium.

Since Sakharov's conditions are satisfied, the baryon asymmetry of the Universe could be created during the electroweak phase transition.  Unfortunately, it has been shown that the amount of CP violation coupled with the strength of a first order phase transition is not sufficient to create enough baryon asymmetry in the SM \cite{Gavela:1993ts, Gavela:1994dt, Huet:1994jb}.  The strength of the phase transition could be boosted by a light (below 80 GeV) Higgs boson but as the Large Hadron Collider confirmed, the mass of the (lightest) standard-like Higgs boson is 125 GeV \cite{Rummukainen:1998as}.

\section{Baryogenesis beyond the Standard Model}

In various extensions of the Standard Model not only Sakharov's conditions can be satisfied but also enough baryon asymmetry can be generated.  The strength of a first order electroweak phase transition can be boosted in supersymmetric extensions by the lightest super-partners of the fermions or by gauge singlet scalar fields.  In (grand) unified theories the phase transition associated with the breaking of the unified gauge group can lead to a sufficiently strong phase transition.  Alternatively, the decay of a new exotic particle, such as a right-handed neutrino, a Q-ball, or the inflaton can be used as a mechanism to deviate from thermal equilibrium.  

A very popular extension of the SM is the Minimal Supersymmetric SM (MSSM).  This model offers several advantages compared to the SM: it contains dark matter candidates, it can explain the existence of a light Higgs boson, it can explain the mechanism of the electroweak symmetry breaking, it can shed light on the different strengths of the standard forces, and it can be connected to gravity.  In the MSSM standard fields are elevated to superfields having both fermionic and bosonic components.  As a result, all standard degrees of freedom receive a super-partner.  The super-partner sector can host new $CP$ violating phases.  Phases associated with the masses of the super-partners of the gauge bosons have been shown to raise the baryon asymmetry \cite{Cline:1996cr, Cline:1997bm, Carena:1997gx, Konstandin:2005cd}.  A caveat is that the same phases also enhance the electric dipole moment (EDM) of the nuclei and leptons, which strongly limits their values above \cite{Cirigliano:2009yd}.  

\begin{figure}[!ht]
	\begin{center}
		\includegraphics[width=0.49\columnwidth]{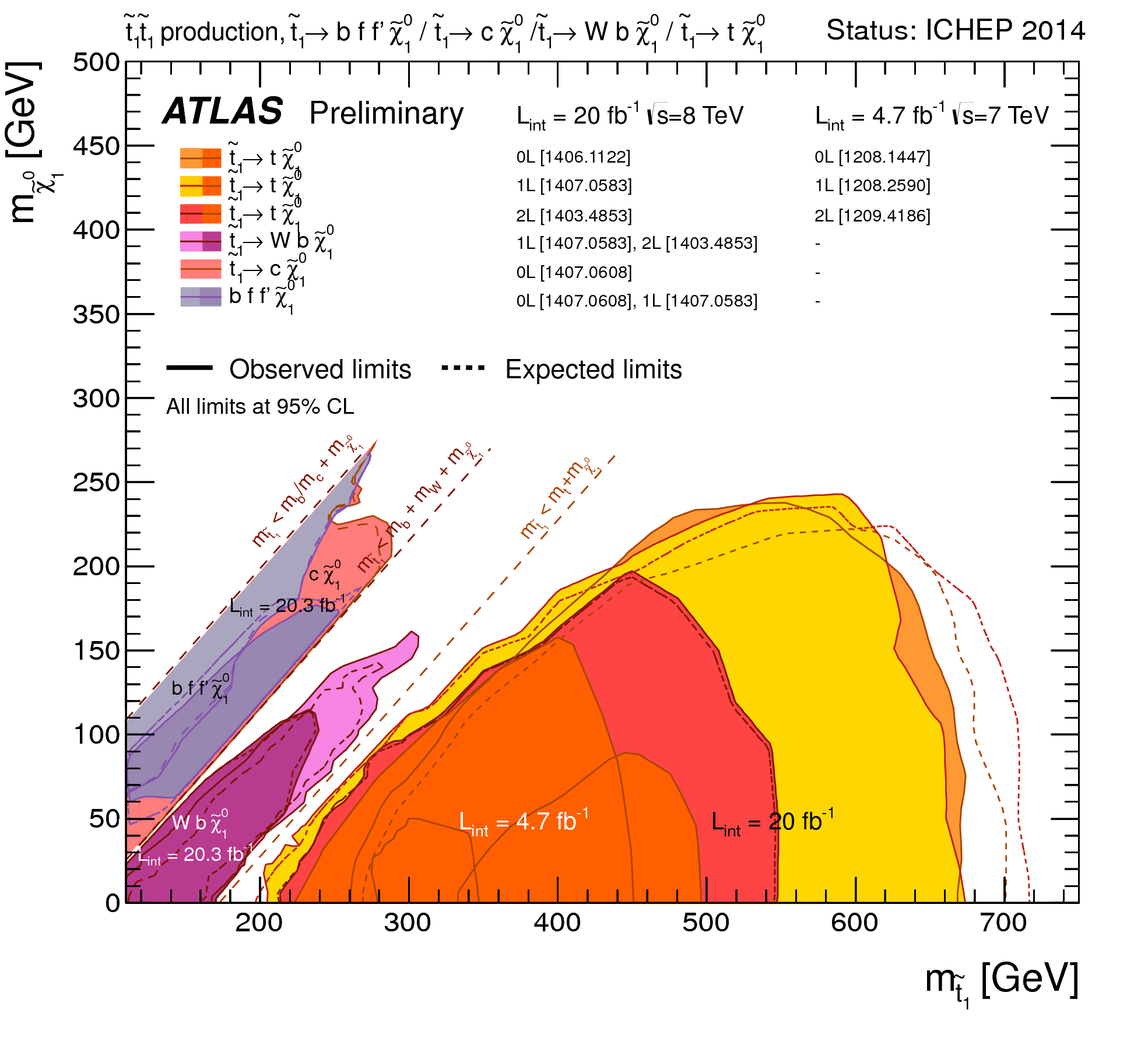}
		\includegraphics[width=0.49\columnwidth]{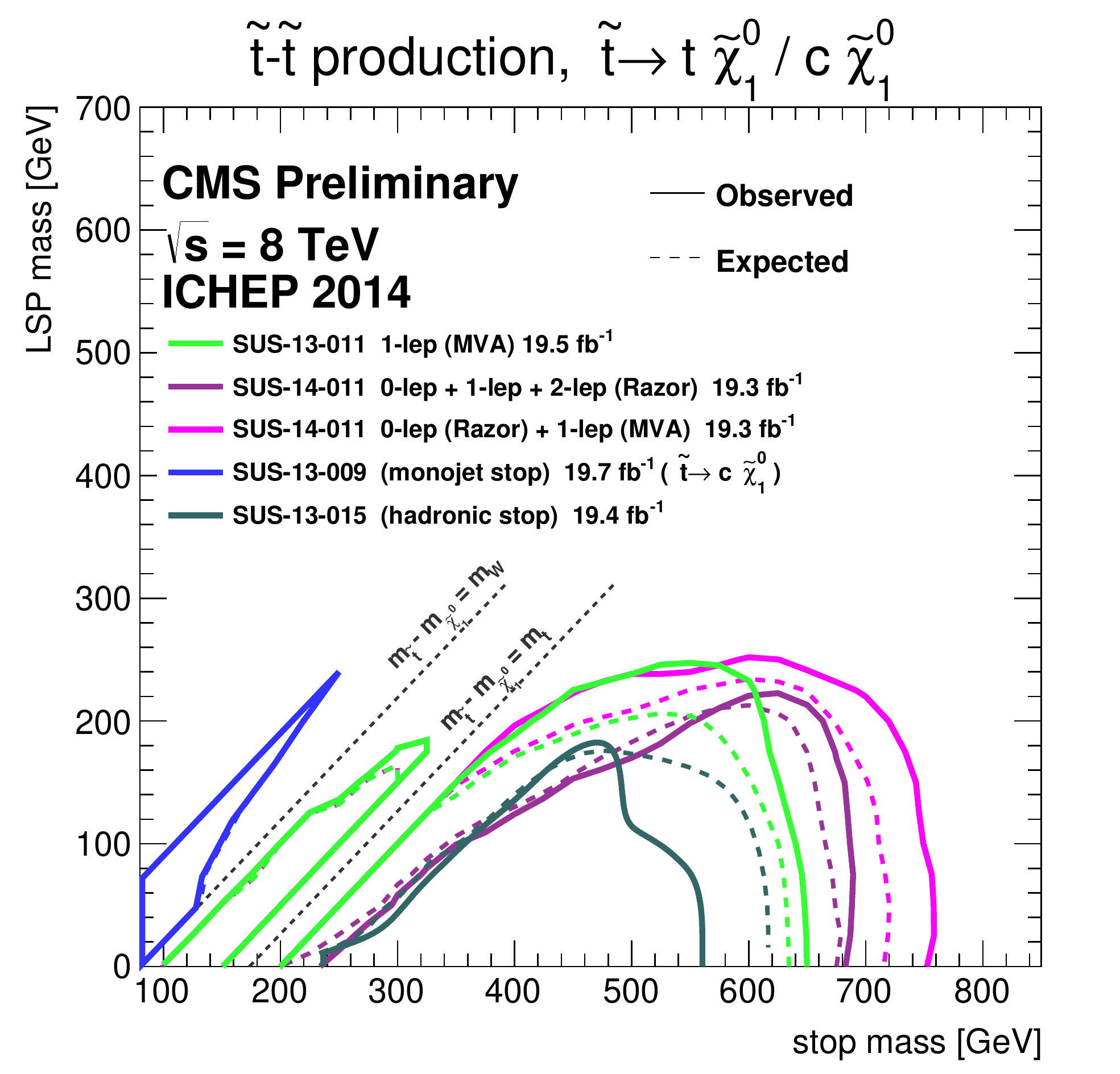}
		\caption{Exclusion regions for stop quarks, the supersymmetric partners of the top quark, set by the ATLAS and CMS collaborations at the Large Hadron Collider.  Narrow parameter regions still remain where stops almost as light as the top quark can hide (based on \cite{Aad:2012ywa, Aad:2012xqa, Aad:2012uu, Aad:2014qaa, Aad:2014bva, Aad:2014kra, Aad:2014nra, CMS-SUS-13-009, CMS-SUS-13-011, CMS-SUS-13-015, CMS-SUS-14-011}).}
		\label{fig:stop}
	\end{center}
\end{figure}

The order parameter of the first order phase transition can also be increased in the MSSM.  This parameter receives contributions from effective three-Higgs vertices.  In the SM electroweak gauge bosons and quarks with large (Yukawa) couplings to the Higgs boson contribute to these vertices.  In the MSSM relatively light superpartners of top and bottom quarks increase the order parameter \cite{Carena:1996wj, Delepine:1996vn, Cirigliano:2006wh}.  This increase is substantial if the mass of these super-partners is below 1 TeV.  Presently, the Large Hadron Collider sets limits on these masses.  These bounds allow narrow regions in the parameter space where electroweak baryogenesis in the MSSM is still viable, as shown in Figure \ref{fig:stop} \cite{Balazs:2004bu, Balazs:2004ae}.  The exploration of these parameter regions is a very important task at Run II of the LHC.

Since the electroweak baryogenesis window in the MSSM has been substantially narrowed by the LHC and recent EDM measurements, considerable attention has been payed to the issue of baryogenesis in models beyond the MSSM.  Perhaps the most obvious model to examine is the Next-to-MSSM.  Compared to the MSSM this model features a gauge singlet superfield.  The scalar component of the singlet couples directly (at tree level) to the Higgs bosons.  This coupling alone realizes a triple Higgs coupling and is capable to generate a high enough order parameter during a first order electroweak phase transition \cite{Balazs:2013cia}.  In this scenario light stops are not needed, thus the LHC stop limits are not constraining.  EDM limits on $CP$ violating phases in the gaugino sector may still pose a challenge, but these phases might be included in the singlet sector to evade the experimental limits.

\section{Summary}

The origin of the cosmic matter-antimatter asymmetry is unknown.  The Standard Model of elementary particle physics satisfies all necessary conditions to generate baryon asymmetry from an initially symmetric phase, however the amount of asymmetry that can be generated falls short of the experimentally observed.  Supersymmetric extensions of the SM fair better.  Especially the NMSSM, which features parameter regions that are allowed by various present experimental limits and where enough asymmetry can be generated.  Near future experimental tests will probe these promising regions.

\bigskip
\section{Acknowledgments}

I thank Marcela Carena, Anupam Mazumdar, Arjun Menon, David E. Morrissey, Michael Ramsey-Musolf, Carlos Wagner, and Graham White who collaborated with me on various projects that involved baryogenesis.  I learned most of what I know about the subject from them.  This work in part was supported by the ARC Centre of Excellence for Particle Physics at the Tera-scale. 




\bibliographystyle{unsrt} 
\bibliography{balazs} 

%
%

%
%
%
%
 
\end{document}